\documentstyle[epsf]{mn} %JO
\newcommand{\etal}{{\em et al.\ }}
\newcommand{\ie}{{\em i.e.\ }}
\newcommand{\cf}{{\em cf.\ }}
\newcommand{\eg}{{\em e.g.\ }}
\newlength{\fsizea}
\newlength{\fsizeb}
\newcommand{\lesssim}{\la} %JO
\newcommand{\gtrsim}{\ga} %JO

\begin{document}

\title[Line-of-sight structure of galaxy clusters] %JO
%PP \title
      {Resolving the line-of-sight structure of galaxy clusters by
       combining X-ray and lensing data}

\author[M.\ Bartelmann \& T.S.\ Kolatt] %JO
%PP \author
       {Matthias Bartelmann$^1$%
        \thanks{e-mail: mbartelmann@mpa-garching.mpg.de} and
        Tsafrir S.\ Kolatt$^{2,3}$%
        \thanks{e-mail: tsafrir@physics.ucsc.edu}\\ 
        $^1$ Max-Planck-Institut f\"ur Astrophysik, P.O.\ Box 1523,
%PP \\
        D--85740 Garching, Germany\\
        $^2$ Harvard-Smithsonian Center for Astrophysics, 
%PP \\
        60 Garden St., Cambridge, MA 02138, U.S.A. \\
        $^3$ The Physics Department and Lick Observatory,
%PP \\
        University of California, Santa Cruz, CA 95064, U.S.A.}

\maketitle

\begin{abstract}
The projected gravitational potential of galaxy clusters is reflected
in both their X--ray emission and their imprint on the images of
background sources due to their gravitational lensing effects. Since
these projections of the potential are weighted differently along the
line-of-sight, we propose a method to combine them and remove the
degeneracy between two cases: (i) a cluster consisting of a single
potential well, or (ii) an apparent cluster composed of several
potential wells projected onto each other. We demonstrate with
simulated data of potential models that this method indeed allows to
significantly distinguish multiple from single clusters. The
confidence limit for this distinction depends on the mass ratio
between the clusters. It ranges from $\sim15\,\sigma$ for mass ratio
1:1 to $\sim4\,\sigma$ for mass ratio 1:6. Furthermore, the method
reconstructs the correct cluster mass, the correct mass ratio of the
two clusters, and the correct scale radii with typical fractional
accuracies of a few percent at $3\,\sigma$ confidence. As an aside,
our method allows to accurately determine gas fractions in clusters,
also with $3\,\sigma$ fractional accuracies of order a few percent.
We argue that our method provides an alternative to the commonly used
$\beta$--fit technique, and yields more reliable results in a broader
range of cases.
\end{abstract}

 %JO

\section{Introduction}
\label{sec:intro}

What physical objects do we call galaxy clusters? Are they typical
large regions of extreme density enhancement? Or do they, as a class,
constitute a sample of peaks in the apparent, line-of-sight projected
density of galaxies, X-ray emission, or dark matter? If so, what do
cluster samples defined by different criteria have in common? Are the
estimates of cluster abundance, and the inferences on cluster
properties, misleading because of projection effects? What are our
chances to identify and quantify projection effects?

The paramount importance of galaxy clusters as probes for the
cosmological evolution of density perturbations and structure
formation makes the answers to these questions essential for any
attempts at interpreting cluster samples.

Despite their undoubted merits, cluster samples compiled by subjective
classification of two-dimensional galaxy-count enhancements (Abell
1958; Zwicky \etal 1968; Abell, Corwin, \& Olowin 1989) are most
susceptible to being statistically unfair. Identifying clusters by
counting galaxies in an automated process (\eg Dalton \etal 1994)
marks a major improvement, but the so-defined samples are still
subject to projection effects.

The prevalence of projection effects is reduced in cluster samples
selected by X-ray surface brightness (\eg Gioia \etal 1990; Ebeling
\etal 1996). Arising mostly from thermal {\em bremsstrahlung\/}, the
X-ray emissivity is proportional to the squared electron density in
the intracluster plasma. It is therefore a much more reliable measure
of the three-dimensional rather than the projected density. Despite
this welcome feature, there is still ample room for selection effects
to be important in some of the analyses based on such samples.

Recently, van Haarlem, Frenk, \& White (1997) demonstrated with
simulations that projection effects are important even for cluster
samples selected by X-ray emission. The line-of-sight integrated X-ray
emission of these clusters is usually fitted with the three-parameter
$\beta$ model (Cavaliere \& Fusco-Femiano 1976). Unfortunately,
conclusions from such fits suffer from projection effects and noise,
and usually cease to provide an adequate functional description of the
dark-matter density profile on intermediate scales of projected radii.

Our ability to recover the line-of-sight (l.o.s.) density structure is
crucial for attempts at constraining cosmological parameters from
cluster samples. It is also vitally important for any assessment of
the physical properties of clusters, \eg the degree of virialization
and of hydrostatic equilibrium of the intracluster gas. For example,
when rich clusters are selected for their strong gravitational lensing
effects (\ie their ability to form large arcs), and then analyzed with
respect to their X-ray data to derive limits on the justification of
the assumption of hydrostatic equilibrium (Miralda-Escud\'e \& Babul
1995; Loeb \& Mao 1994), it may well be that selection effects play an
important role, and that some of the conclusions can be relaxed by
taking projection effects into account (Bartelmann \& Steinmetz 1996).

Another important application concerns the use of the
Sunyaev--Zel'dovich effect (\eg Sunyaev \& Zel'dovich 1980; Rephaeli
1995). The effect can be used in tow alternative ways.  First, we can
assume the line-of-sight extent of the cluster gaseous component
(e.g. by relating it to the angular size using a Hubble constant).
Then, by examining the distortion of the CMB spectrum, we can get
limits for the gas content of the cluster and its temperature. On the
other hand, we can assume the latter two (or estimate them
differently) and deduce the Hubble constant by the comparison of the
angular and line-of-sight extent. In either case, the result depends
strongly on the true l.o.s.\ gas profile. If the latter is not well
known, neither the Hubble constant nor the gas content can reliably be
determined (see, \eg, Roettiger \etal 1997; Holzapfel \etal 1997).

Turning to clusters as tracers of the large-scale structure, we are
facing the same problem again. Knowledge of the l.o.s.\ cluster
profile is important for attempts at deriving the cluster abundance
(White, Efstathiou, \& Frenk 1993; Eke, Cole, \& Frenk 1996; Viana \&
Liddle 1996), the cluster mass function (\eg Bahcall \& Cen 1993;
Burns \etal 1996), the spatial distribution of clusters (\ie
correlations, probability distributions etc., see Bahcall 1988 for a
review), and the cluster velocity dispersion (\eg Fadda \etal 1996;
Mazure \etal 1996).

In this paper, we propose an alternative to the traditional
$\beta$--fit analysis and argue that, at least to some extent,
degeneracies due to projection effects can be broken. The proposed
alternative rests on a simple idea. In hydrostatic equilibrium, it
must be possible to describe with the same gravitational potential all
observable X-ray and lensing data pertaining to a given cluster. The
specific relation between the potential and the X-ray emission depends
on the equation of state and the temperature structure of the gas. It
is sufficient to fix these two (and the connection between the dark
matter and spatial gas distribution) in order for the method to
work. Other assumptions do not affect the principle of our approach.

The available observational data are (i) the line-of-sight integrated
X-ray flux, (ii) the emission-weighted gas temperature, and (iii) the
gravitational lensing effects of the cluster that give rise to, \eg,
coherent distortions of the images of background sources. It is
especially the combination of X-ray and lensing measurements that
promises to break the degeneracies arising from projection effects. In
particular, the X-ray flux is most sensitive to the gas fraction and
the physical extent of the system along the line-of-sight. The
emission-weighted temperature is most sensitive to the depth of the
three-dimensional potential well, and the shear field is most
sensitive to the integrated gravitational potential (with the nice
feature of being indifferent to the gas content).

For the sake of demonstration, we investigate two classes of
three-dimensional potentials: (i) a single, isolated cluster, and (ii)
two well separated clusters projected onto each other along the
line-of-sight. In both cases, we describe the intracluster plasma as
an isothermal ideal gas with spatially constant mean molecular weight.
Fixing the functional form of the gravitational potential, we simulate
``observed'' X-ray flux maps, X-ray spectra, and lensing distortion
maps. Given these synthetic observations, we search the parameter
space of this functional form. We minimize an appropriate $\chi^2$
function which contains contributions from all three types of data. As
we shall show, the best-fit model parameters reproduce the input
potential very reliably. Moreover, the degeneracy between the one- and
two-cluster solutions is removed. Attempts to fit a functional form
different from the one simulated result in a very poor goodness-of-fit
relative to the goodness-of-fit for the correct functional form.

We start (\S\ref{sec:obs}) by specifying the explicit and implicit
assumptions we make and provide a concise description of the
observations. We proceed by relating the observables to the underlying
gravitational potential (\S\ref{sec:obs}). In the same section, we
present the functional form of the potential we choose. The
combination of all observables, and the models to describe them,
allows us to write down a $\chi^2$ statistic (\S\ref{sec:chi}), which
we minimize in order to find the most probable solution for a set of
data in the framework of a specific model. In \S \ref{sec:simu}, we
simulate observations of clusters with a specific density profile by
mimicking real observations along with their errors, and demonstrate
the ability to recover the correct (input) density profile from the
projected quantities. We then try to fit a wrong model for the
simulated system and show how we fail in doing so.  In
\S\ref{sec:comp_beta}, we present the difficulties of the $\beta$-fit
model to recover the right cluster parameters.  In \S\ref{sec:conc} we
discuss future implications of this method and present our
conclusions.  We consider this paper as a simple, but necessary, first
step towards lifting the line-of-sight degeneracy in galaxy
clusters. We present the basic ideas here, postponing more detailed
studies to later work.

\section{Model Potential and Observables}
\label{sec:obs}

We employ spherically symmetric cluster models and assume that the
X-ray emitting intracluster gas is in hydrostatic equilibrium with the
cluster gravitational potential. We assume that the gas is isothermal,
and the mean molecular weight is constant throughout the cluster.
These assumptions determine the density profile of the gas. In order
to normalize the gas density, we fix the ratio between gas mass and
dark mass within a given radius.

Then, specifying the three-dimensional cluster potential is sufficient
to describe both the X-ray emission and the lensing properties of the
cluster. Unfortunately, we do not know this three-dimensional
potential {\em a priori\/}. Rather, we must elect a functional form
for it, based on some additional information, which can for instance
be taken from numerical simulations.

There is growing evidence that the averaged radial structure of
numerically simulated dark halos can well be described by a universal,
two-parameter family of density profiles $\rho(r)$,
\begin{equation}
  \rho(r) = \frac{\rho_{\rm s}}{x(1+x)^2}\;,
\label{eq:1}
\end{equation}
where $x$ is the radius in units of a scale radius $r_{\rm s}$. This
shape of the density profile results independently of the parameters
of the background cosmological model, and for halos with a broad range
of masses (Navarro, Frenk, \& White 1996; Cole \& Lacey 1996; Huss,
Jain, \& Steinmetz 1997). On the observational side, Carlberg \etal
(1997) have shown that the velocity dispersion profiles of observed
clusters are compatible with density profiles of the form
(\ref{eq:1}). If the X-ray emitting gas is isothermal and in
hydrostatic equilibrium with the dark-matter profile (\ref{eq:1}), its
flux profile has a flat core despite the cusp in the density profile.
Density profiles with small core radii or central singularities better
fit the observations of giant arcs. The latter require cluster density
profiles with much smaller cores than inferred from X-ray observations
(see \eg Narayan \& Bartelmann 1997 for a review).

If the gravitational potential of the density distribution
(\ref{eq:1}) is normalized such that $\Phi\to0$ for $x\to\infty$, it
can be written
\begin{equation}
  \Phi(r) = -4\pi G\rho_{\rm s}\,r_{\rm s}^2\,\frac{\ln(1+x)}{x}\;.
\label{eq:2}
\end{equation}
We replace the parameter $\rho_{\rm s}$ by the virial mass, by which
we mean the mass contained within the radius $r_{200}$ which encloses
an average overdensity of $\delta_{\rm c}=200\;$\footnote{This choice
of $\delta_{\rm c}$ can be viewed as merely a change of variables. The
actual value for the overdensity within the virialized region may
change as function of the background cosmology, and is of no
particular importance here.}.  Since the mass within radius $r$ is
\begin{equation}
  M(r) = 4\pi\rho_{\rm s}r_{\rm s}^3\,
  \left[\ln(1+x) - \frac{x}{1+x}\right]\;,
\label{eq:3}
\end{equation}
$r_{200}$ is determined by
\begin{equation}
  \frac{3\,\rho_{\rm s}}{c^3}\,
  \left[\ln(1+c) - \frac{c}{1+c}\right]\,
  = 200\,\bar\rho\;,
\label{eq:4}
\end{equation}
where $\bar\rho$ is the mean cosmic density, and $c=r_{200}\,r_{\rm
s}^{-1}$ is a concentration parameter.

An alternative source of information for the functional form of the
potential could be derived from the observed projected X-ray flux
profile. This leads to the famous derivation of the $\beta$-fit. We
discuss the comparison between the adequacy of the two different
functional forms later in \S\ref{sec:comp_beta}.

\subsection{X--ray Emission}
\label{subsec:obs_x}

An isothermal gas in hydrostatic equilibrium with a potential $\Phi$
has a gas density of
\begin{equation}
  \rho_{\rm gas}(r) = \rho_{{\rm gas},0}\,\exp\left[
  -(\tilde\Phi(r)-\tilde\Phi_0)\right]\;,
\label{eq:5}
\end{equation}
where the index `0' refers to some fiducial radius $r_0$. $\tilde\Phi$
is the potential in units of the square of a fiducial velocity $v_{\rm
th}$ of the gas particles,
\begin{equation}
  \tilde\Phi(r) = v_{\rm th}^{-2}\,\Phi(r) = 
  \frac{\bar m}{kT}\,\Phi(r)\;,
\label{eq:6}
\end{equation}
$\bar m$ being the mean mass per particle. For a mixture of $75\%$
hydrogen and $25\%$ helium (by mass), which we henceforth adopt, $\bar
m\simeq10^{-24}\,{\rm g}$. For $r_0$, we choose the virial radius,
$r_{200}$. We further adapt $\rho_{{\rm gas},0}$ such that the total
gas mass within the virial radius is a fraction $f_{\rm gas}$
(hereafter called ``gas fraction'') of the total mass,
\begin{eqnarray}
  4\pi\,\int_0^{r_{200}}\,r^2\,{\rm d}r\,\rho_{\rm gas}(r) &=&
  f_{\rm gas}\,M(r_{200}) \nonumber\\
  &=& f_{\rm gas}\,\frac{4\pi}{3}\,r_{200}^3\,200\bar\rho\;,
\label{eq:6a}
\end{eqnarray}
where the last equality follows from the definition of $r_{200}$.

The emissivity of the gas due to thermal {\em bremsstrahlung\/} at
position $\vec x$ in the energy range $E_{\rm a}\le E\le E_{\rm b}$ is
\begin{eqnarray}
  j_{\rm X}(\vec x;E_{\rm a},E_{\rm b}) &=&
  5.53\times10^{-24}\,{\rm erg\,cm^{-3}\,s^{-1}}\,\nonumber\\
  &\times&
  \left(\frac{kT}{{\rm keV}}\right)^{1/2}\,
  \left(\frac{n_{\rm e}}{{\rm cm}^{-3}}\right)^2\,\nonumber\\
  &\times&
  \left[\exp\left(-\frac{E_{\rm a}}{kT}\right) -
        \exp\left(-\frac{E_{\rm b}}{kT}\right)\right]\;.
\label{eq:7}
\end{eqnarray}
Assuming complete ionization, the electron density is $n_{\rm
e}=0.52\,\bar m^{-1}\rho_{\rm gas}$. The flux $S_{\rm X}(\vec\xi)$
received from the two-dimensional position $\vec\xi$ within the
cluster is the line-of-sight integral
\begin{equation}
  S_{\rm X}(\vec\xi;E_{\rm a},E_{\rm b}) =
  \frac{1}{4\pi\,(1+z)^3}\,\int\,{\rm d}l\,
  j_{\rm X}(\vec\xi,l;E'_{\rm a},E'_{\rm b})\;,
\label{eq:7a}
\end{equation}
where the factor $(1+z)^3$ accounts for redshifting the photons and
for the ratio between luminosity distance and angular-diameter
distance, and $E'_{\rm a,b}=(1+z)E_{\rm a,b}$.  In real observations,
$S_{\rm X}(\vec\xi;E_{\rm a},E_{\rm b})$ is further convolved with the
detector response function.  The flux can then be converted to photon
numbers by means of the {\em bremsstrahlung\/} spectrum, the detector
area, and the exposure time. Likewise, the observed photon spectrum is
determined by the number of photons per energy bin $[E_i,E_{i+1}]$,
$E_{\rm a}\le E_i\le E_{\rm b}$.

\subsection{Gravitational Lensing}
\label{subsec:obs_e}

The gravitational lensing effects of the density profile (\ref{eq:1})
have been calculated elsewhere (Bartelmann 1996). Given the potential
(\ref{eq:2}), the effective lensing potential is
\begin{equation}
  \psi = \frac{D_{\rm ds}}{D_{\rm d}D_{\rm s}}\,
  \frac{2}{c^2}\,\int\,{\rm d}l\,\Phi\;,
\label{eq:8}
\end{equation}
where $D_{\rm d}$, $D_{\rm s}$, and $D_{\rm ds}$ are the
angular-diameter distances from the observer to the cluster, to the
sources, and from the cluster to the sources, respectively. The
lensing convergence $\kappa$ and shear components $\gamma_{1,2}$ are
then
\begin{eqnarray}
  \kappa &=& \frac{1}{2}(\psi,_{11}+\psi,_{22}) \nonumber\\
  \gamma_1 &=& \frac{1}{2}(\psi,_{11}-\psi,_{22}) \quad,\quad
  \gamma_2 = \psi,_{12}
\label{eq:9}
\end{eqnarray}
where indices $i$ preceded by a comma denote partial derivatives with
respect to $x_i$. The lensing properties of mass profiles of the form
(\ref{eq:1}) have been worked out by Bartelmann (1996).

Gravitational lensing leads to coherent distortions of the images of
background galaxies. Image ellipticities, which can be quantified by,
e.g., the quadrupole tensor of their surface brightness distribution,
measure the two-component reduced shear
\begin{equation}
  g_i = \frac{\gamma_i}{1-\kappa}\;.
\label{eq:10}
\end{equation}
If the lens has a critical curve, an ambiguity arises in the $g_i$
because of the parity change upon crossing the critical curve. An
unambiguous measure of the ellipticity is then provided by the
distortion $\delta_i$,
\begin{equation}
  \delta_i = \frac{2\,g_i}{1+g_1^2+g_2^2}\;.
\label{eq:11}
\end{equation}

Since galaxies do not usually appear circular, $\delta$ cannot be
inferred from individual galaxies, but must be determined
statistically by averaging over a sufficient number of galaxy
images. The assumption underlying this inference is that the intrinsic
{\em orientations\/} of the galaxies is random. The measured galaxy
ellipticities (to be related to the $g_i$) are given by
\begin{equation}
  \epsilon_1+{\rm i}\,\epsilon_2 = \frac{a-b}{a+b}\,
  \exp(2\,{\rm i}\,\varphi)\,
\label{eq:ellip_def}
\end{equation}
with $a$ and $b$ the major and minor axes of the ellipse,
respectively, and $\varphi$ its orientation (position angle). The
unlensed observed ellipticities follow the two-dimensional
distribution
\begin{equation}
  p_{\rm e}(\epsilon_1,\epsilon_2) =
  \frac{\exp(-|\epsilon|^2\,\sigma_\epsilon^{-2})}
       {\pi\sigma_\epsilon^2\,[1-\exp(-\sigma_\epsilon^{-2})]}\;,
\label{eq:12}
\end{equation}
with $\sigma_\epsilon\sim0.15$ (\eg Miralda-Escud\'e 1991; Tyson \&
Seitzer 1988; Brainerd, Blandford, \& Smail 1996). An iterative
procedure to derive $\delta$ from galaxy ellipticities has been
described by Seitz \& Schneider (1995).

Deep observations (\eg Smail \etal 1995) find galaxy surface number
densities of $\sim40-50\,{\rm arcmin}^{-2}$ down to a magnitude limit
of $R\sim25$. According to Lilly \etal (1995), the average redshifts
of such sources fall within $0.8-1$. If we want to average over
$\sim10$ galaxies for each local estimate of the distortion, the
intrinsic resolution limit for any such distortion map is
$\sim30''$. The uncertainty in the local determination of $\delta$ can
be estimated by the variance of the $N'$ galaxy ellipticities used to
determine $\delta$, divided by $(N'-1)^{1/2}$.

\section{Combined $\chi^2$ Function}
\label{sec:chi}

What question can we answer by calling statistics to our assistance?
We do not know {\em a priori\/} whether a certain model provides a
good description to the data. We can, however, find answers to the
three following questions:

\begin{enumerate}

\item Given the data, what are the best parameters to describe them,
{\em in the framework of a specific model\/}?

\item For these best parameters found earlier, how likely is the
model, given the data ?

\item Given the data, which of $n$ competing models is the most
likely?

\end{enumerate}

The first answer is provided by the $\chi^2$ minimization, the second
by the goodness-of-fit (GoF) evaluation, and the third by comparing
GoF values as obtained for the $n$ different models. By ``model'' we
mean a functional parameterization of the three-dimensional
gravitational potential of the system under consideration.

The ability to answer the aforementioned three questions is pending on
our ability to constitute a decent $\chi^2$ statistic. The $\chi^2$
statistic can be easily interpreted if the error estimate is accurate,
and the error distribution is Gaussian or close to Gaussian.

Our analysis makes use of the different sensitivity of the observables
to the potential parameters. The $\chi^2$ statistic should therefore
take into account all observables simultaneously. For the clarity of
presentation, however, we present the various terms in the statistic
separately and combine them later.

\subsection{The Temperature Term: $\chi^2_{\rm T}$}
\label{subsec:chi2_T}

The first term in the $\chi^2$ statistic deals with the
emission-weighted temperature (``temperature'' hereafter). In
\S\ref{subsec:obs_x} we described the photon counts in the $N_{\rm E}$
photon energy bins from which the temperature is estimated. The
assumption of cylindrical symmetry, the independence of temperature on
projected radius due to the assumption of isothermality, and the poor
spatial resolution of the observations lead us to consider only one
annulus, centered on the X--ray flux centroid, for the temperature
evaluation.

The overall number of photons is taken into account elsewhere (the
flux term, see \S\ref{subsec:chi2_S}), and we must avoid taking it
into account twice, or otherwise the terms would not be
independent. We therefore normalize the photon number in each energy
bin by the total observed number of photons in all energy bins.

In normalized units, the temperature term of the $\chi^2$ statistic is
\begin{equation}
\label{eq:chi2_t}
  \chi^2_{\rm T} = \sum_{i=1}^{N_{\rm E}}\,
  \frac{[n_\gamma^i-n_\gamma(E_i)]^2}{(\sigma^i_{\rm T})^2}\;.
\end{equation}
The normalized photon count in energy bin $E_i$ is $n_\gamma^i$. Given
the model temperature and assuming {\em bremsstrahlung\/} radiation,
the model for the cluster and the X-ray background predicts
$n_\gamma(E_i)$ photons in the same bin \footnote{When more than one
annulus is taken into account, the l.o.s.\ integration with the
relevant emission weighting must be carried out as well and added to
$\chi^2_{\rm T}$.}. The error in the denominator has two contributions
that add up in quadrature: $\sigma_{\rm T}^2=\sigma_{{\rm
T},1}^2+\sigma_{{\rm T},2}^2$. The two kinds of measurement error are
the instrumental error and the background radiation that has to be
estimated in each frequency bin. These errors are identical for all
models, since we do not consider different X--ray background radiation
models. We assume the average X-ray background radiation signal is
known, so there is no explicit DC offset of the photon number
counts. Fluctuations in the background, though, should still be taken
into account. A gross approximation for the contributions of these two
terms can be the square root of the number of observed photons, if the
observational errors are all due to Poisson noise. We thus have
\begin{equation}
\label{eq:sig_T_12}
  (\sigma^i_{{\rm T},1})^2 + (\sigma^i_{{\rm T},2})^2
  \simeq n_\gamma^i\;.
\end{equation}
Recall that here, too, we normalize by the overall number of {\em
observed\/} photons. One has to stay away from very low photon number
counts, where the result is biased by the lower limit of detecting no
photons at all, and the error symmetry breaks down.

If a model assumes more than one cluster, with different temperatures,
one cannot avoid the integration needed to calculate the weighted sum
that yields the expected projected temperature.

Note that longer integration time can reduce the relative importance
of $\sigma_{\rm T,1}$, but cannot help reduce the relative importance
of the background radiation noise ($\sigma_{\rm T,2}$) as long as it
is not due to a short temporal variation.

\subsection{The Flux Term: $\chi^2_{\rm S}$}
\label{subsec:chi2_S}

The $\chi^2$ term for the flux is similar to the term for the
temperature. In both cases the data is the number of photons. Two
distinct differences exist between the two: In the flux term, the
independent data are numbers of photons in spatial pixels, and the
important measure is the actual number, so it can not be normalized by
the total number observed.

For $N_{\rm p}$ pixels, with the $i^{\rm th}$ pixel centered on
$\vec\xi_i$, $N_\gamma^i$ measured photons in the pixel, and
$N_\gamma(\vec\xi_i)$ photons expected from the model, the flux
$\chi^2$ term is written as
\begin{equation}
  \chi^2_{\rm S} = \sum_{i=1}^{N_{\rm p}}\,
  \frac{[N_\gamma^i-N_\gamma(\vec\xi_i)]^2}{(\sigma^i_{\rm S})^2}\;.
\label{eq:chi2_S}
\end{equation}
Similarly to the $\sigma_{\rm T}$ calculation, here too we have the
same two contributions which we can approximate by $(\sigma^i_{\rm
S})^2 \simeq N_\gamma^i$. In the case of the flux, we are interested
in both the absolute number of photons and their spatial distribution
in the projected two-dimensional map as a function of $\vec\xi_i$.

\subsection{The Ellipticity Term: $\chi^2_\delta$}
\label{subsec:chi2_E}

The data ``unit'' for the shear field, as explained in
\S\ref{subsec:obs_e}, is an area of typically $0.2\,{\rm arcmin}^2$ in
which there are enough background galaxies ($\sim10$) to average over,
for deriving the mean reduced distortion (\cf eq.~\ref{eq:11}) in this
area element ($\langle\vec\delta\rangle$). The error in the derived
distortion in each bin is model independent and can be calculated
either by the dispersion about the average distortion in a given area,
or by taking a non-lensed region, deriving the intrinsic ellipticity
distribution for the same galaxy population, and dividing by the
square root of the number of galaxies in each bin. The two methods
give similar results of $\sigma_\delta\sim0.03$. Since the area
element sizes are identical across the cluster, so is the error for
the average ellipticity values. The $\chi^2$ term for the
ellipticities is readily written
\begin{equation}
\label{eq:chi_E}
  \chi^2_\delta = \sum_{i=1}^{N_\delta}\, \frac{[\langle\vec
  \delta\rangle_i- \langle\vec\delta(\vec\xi_i)\rangle]^2}
  {\sigma_\delta^2}\;,
\end{equation}
where, as usual, $\langle\vec\delta(\vec\xi_i)\rangle$ is the
distortion expected from the model about the position $\vec\xi_i$. The
sum is over all regions, $i$, for which the ellipticity is evaluated
($N_\delta$ regions altogether).

\subsection{The Combined $\chi^2$}
\label{subsec:chi}

As stated earlier, the idea is to search for $\chi^2$ minima in the
potential parameter space, using all observables simultaneously. Say
we have specified a functional form for the potential $\Phi(\vec r)$
that involves only two fitting parameters, plus one parameter for the
gas fraction. The X-ray temperature, the X-ray flux, and the
distortion field in any bin, are all functionals of this
potential. They can all be combined to result in one (complicated,
non-linear) function. So can the data be combined. The overall
$\chi^2$ statistic has therefore $N_{\rm E}+N_{\rm p}+N_\delta-3$
degrees of freedom (d.o.f.), and is simply the sum
\begin{equation}
\label{chi}
  \chi^2 = \chi^2_{\rm T}+\chi^2_{\rm S}+\chi^2_\delta\;.
\end{equation}
Notice that the number of d.o.f.\ for this function is {\em not\/} the
sum of the number of d.o.f.\ when each individual term is considered
separately.

Different models may have different numbers of fitting parameters
[e.g. two spherical symmetric clusters along the l.o.s.\ may be
specified by five or six (depending on the universality of $f_{\rm
gas}$) plus the separation between the two clusters]. A fair
comparison between models must include this simple fact.

The $\chi^2$ minimization leaves us with a $\chi^2_{\rm min}$ for each
model, and an estimate of the best fitting parameters for this
model. In order to assess to what extent a specific model provides an
adequate description for the data, the GoF is calculated according to
\begin{equation}
\label{eq:gof}
  G = \Gamma\left(\frac{N_{\rm dof}}{2},
                  \frac{\chi^2_{\rm min}}{2}\right)\;,
\end{equation}
with $\Gamma$ the incomplete gamma function. The GoF interpretation
rests on two assumptions: (i) that the data that went into the
$\chi^2$ calculation are independent (so that the d.o.f.\ calculation
truly represents the d.o.f.\ of the data and the model), and (ii) that
the errors are distributed in a Gaussian fashion and
uncorrelated. Validation of the second assumption can be carried out
by inspection of the residuals distribution. If errors are indeed all
due to a Poissonian process in the data collection, we have reasons to
believe that by the central limit theorem, the errors are
Gaussian. The prudent policy of taking large bins (in the relevant
context for each observable), pays off by producing minimally
correlated errors and independent processed data points.

\section{Demonstration by Simulations}
\label{sec:simu}

\subsection{Model Specification}

\begin{figure*} %JO
%PP \begin{figure}[ht]
\mbox{\epsfxsize=\fsizea\epsffile{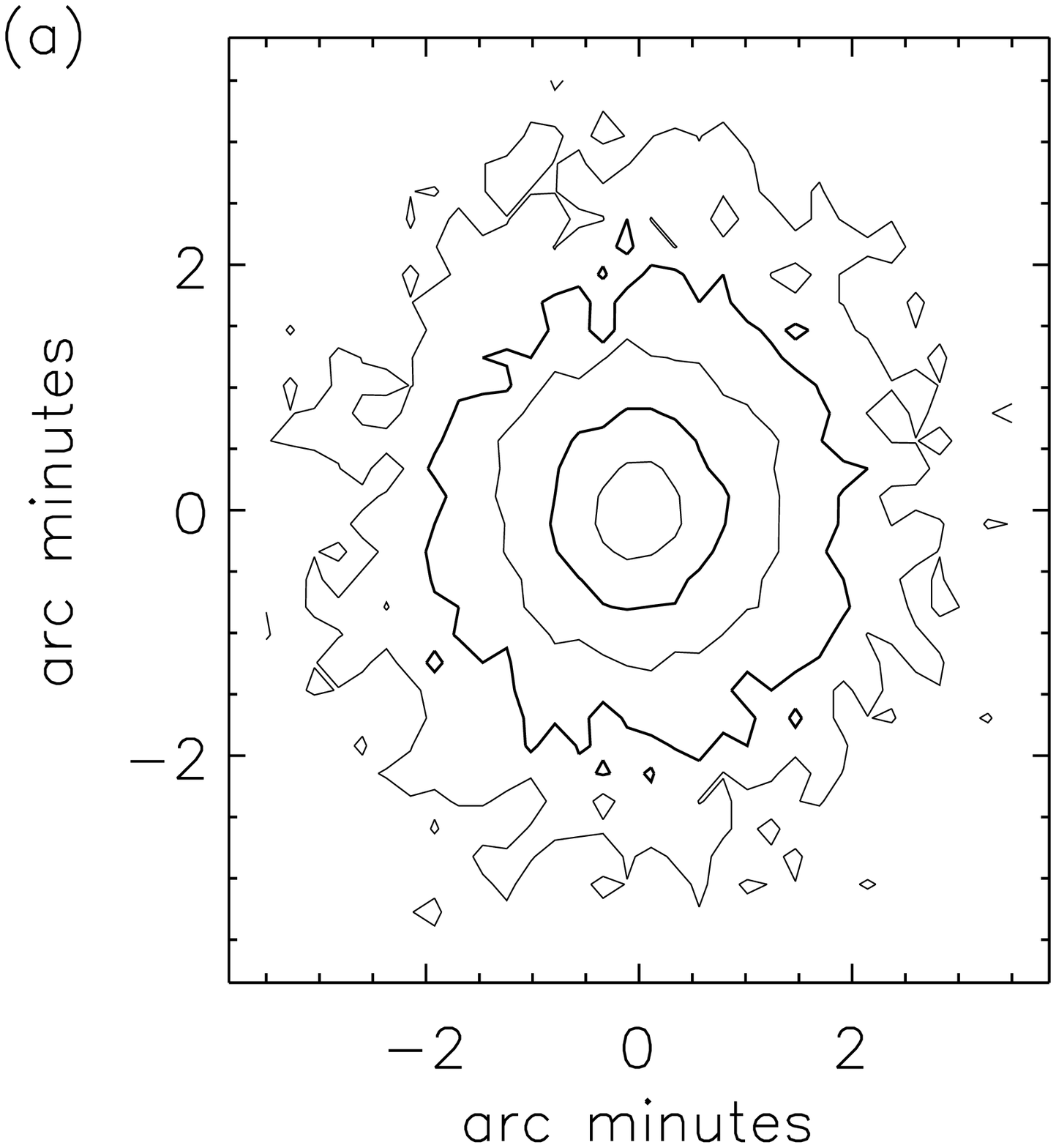}
      \epsfxsize=\fsizea\epsffile{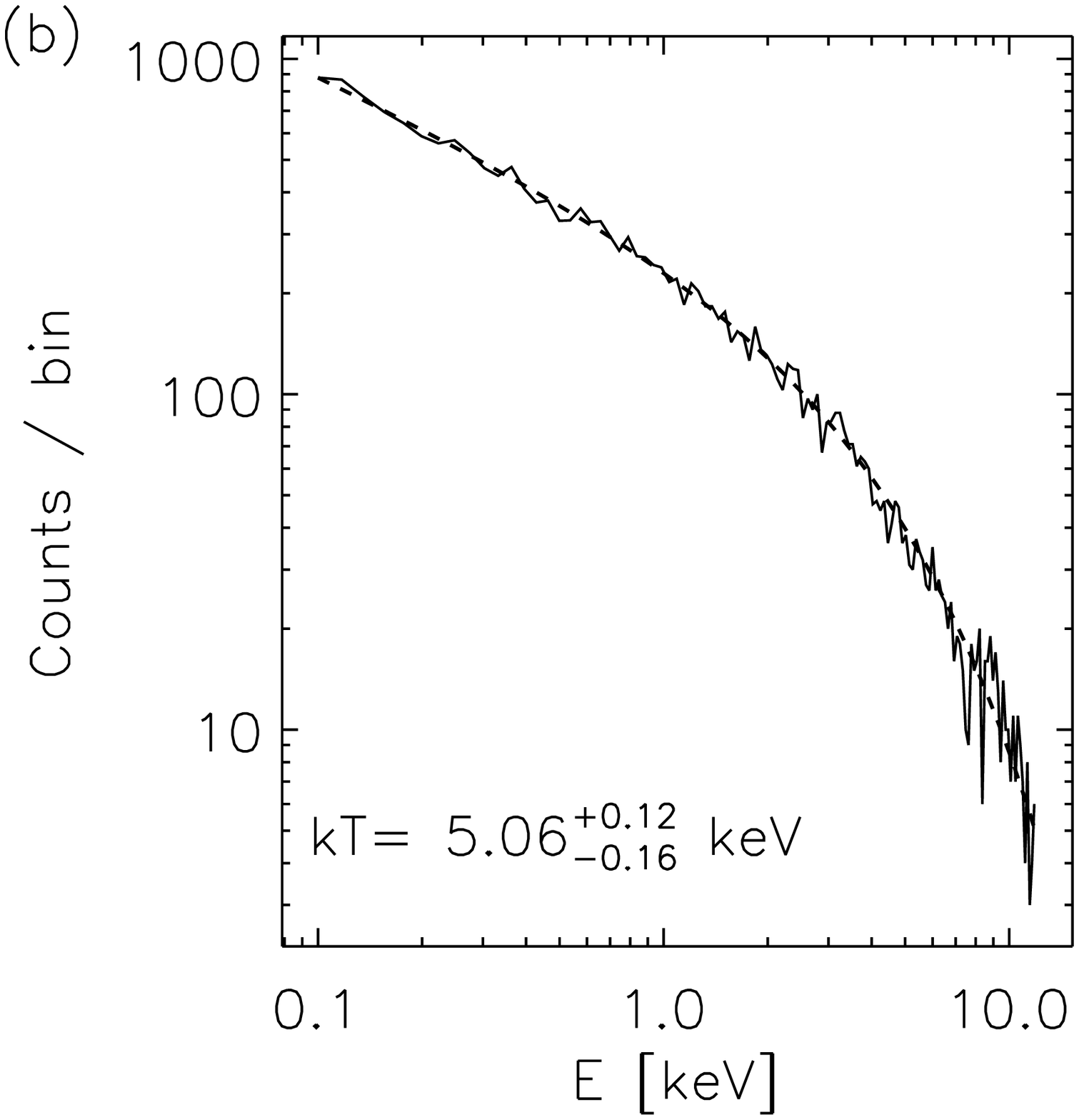}
      \epsfxsize=\fsizea\epsffile{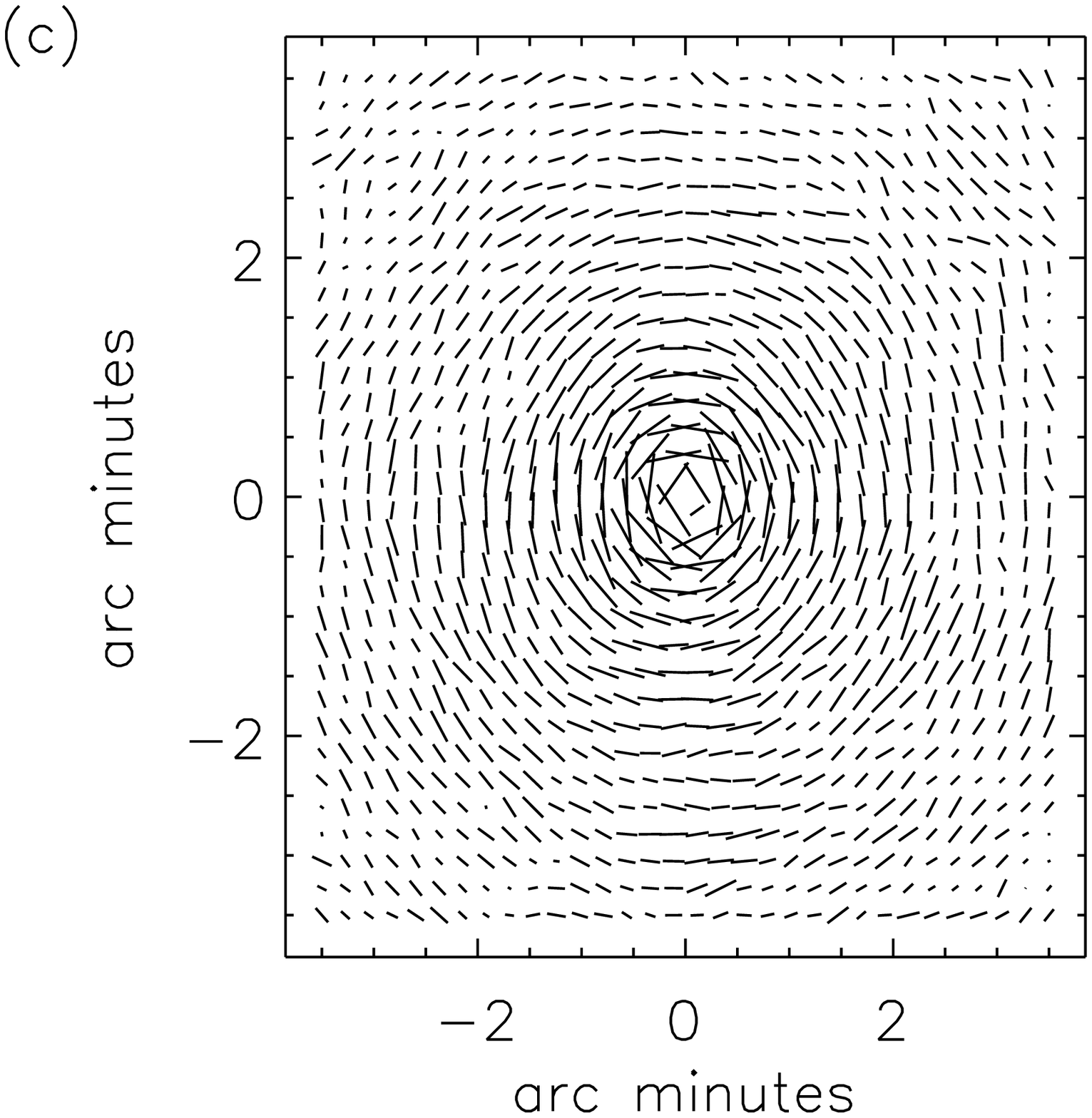}}
\caption{Examples for the simulated data we use. Panel (a): Simulated
  X-ray flux map. The contours are spaced by a factor of $10^{1/2}$ in
  units of counts per pixel. The pixel size is $13''\times13''$, the
  field size is $7'\times7'$ (the field has $32\times32$
  pixels). Panel (b): Simulated X-ray spectrum, overlaid with a fit to
  the {\em bremsstrahlung\/} spectrum (dashed curve). The best-fit
  temperature is given in the plot, together with its $1\,\sigma$
  error. Panel (c): Distortion map produced from simulated background
  galaxy ellipticities. The same potential was used as for the X-ray
  data in the other panels. The length of the lines indicates the
  modulus of $\delta$, their orientation shows the direction of
  $\delta$.}
\label{fig:1}
\end{figure*} %JO
%PP \end{figure}

We can now proceed and apply our technique to idealized test cases. We
consider two such cases: either one isolated cluster, or two clusters
projected onto each other along the line-of-sight. In the first case,
all observables are completely specified by three parameters, viz.\
the two parameters of the dark-matter profile, for which we take the
virial mass $M_{\rm vir}$ and the scale radius $r_{\rm s}$, and the
gas fraction, $f_{\rm gas}$ (by weight) within the virial radius.

Assuming that the gas fraction is ``universal'', we have five
parameters to describe two clusters, plus their mutual distance. If
the clusters are sufficiently distant such that their gas
distributions do not significantly overlap, their exact separation
does not matter. This applies once they are separated by more than
about the sum of their scale radii. If they are so close to each other
that they are in the process of merging, their gas distributions
become more complicated, especially because shocks form, hydrostatic
equilibrium does not apply, and their dark-matter distributions are
deformed. We assume here that the two clusters are sufficiently well
separated such that their gas distributions do not interact, and then
five parameters suffice to characterize their X--ray and lensing
properties.

We consider simulated clusters at a redshift of $z_{\rm c}=0.2$, with
a gas fraction of $f_{\rm gas}=10\%$, scale radii of $r_{\rm
s}=0.25\,h^{-1}\,$Mpc, and a total virial mass of $M_{\rm
vir}=10^{15}\,M_\odot$. When there are two clusters, this is the sum
of the individual virial masses. The lensed sources are put at a
redshift of $z_{\rm s}=1$.

We choose energy bins such as to mimic the energy resolution of the
ASCA SIS (Tanaka, Inoue, \& Holt 1994),
\begin{equation}
  \frac{\Delta E}{E} = 0.02\,
  \left(\frac{E}{5.9\,{\rm keV}}\right)^{-1/2}\;,
\label{eq:7b}
\end{equation}
which results in $N_{\rm E}=121$ energy bins between $E_{\rm
a}=0.1\,$keV and $E_{\rm b}=12\,$keV. The energy dependence of the
effective detector area is modeled like that of the ROSAT HRI. The
spectral energy distribution yields the emission-weighted temperature,
which equals $T$ for isothermal gas in a single cluster, or in a
double cluster where both components have equal mass. For the noise in
the photon counts, we use Poisson noise plus an additional background,
for which we choose a DC level of $3\times10^{-4}\,{\rm
s^{-1}\,arcmin^{-2}}$, and a Gaussian distribution of the variation
about this level with the standard deviation of $\sigma_{{\rm
S},2}=N_{\rm b}^{1/2}$, with $N_{\rm b}$ the number of background
photons per exposure. This is in approximate agreement with the
background noise in the ROSAT PSPC (Snowden \etal 1995). We ignore any
energy dependence of the X-ray background. An example for a flux map
and a spectrum simulated this way is shown in panels (a) and (b) of
Fig.~\ref{fig:1}. Throughout, we have assumed an exposure time of
$10\;{\rm ksec}$.

For the background galaxies, we choose a surface number density of
$40\,{\rm arcmin}^{-2}$ at a redshift of $z_{\rm s}=1$. Their
positions are random, and their unlensed ellipticities are drawn from
the two-dimensional distribution of eq.~(\ref{eq:12}) with
$\sigma_\epsilon=0.15$. The galaxies are then distorted by the lensing
effect of the simulated clusters, and the distortion $\delta$ is
determined using the iterative algorithm provided by Schneider \&
Seitz (1995).  An example for the distortion map $\delta$ created by
the lensing effect of the cluster whose X--ray emission is shown in
panels (a) and (b) of Fig.~\ref{fig:1}, is displayed in panel (c) of
the same figure.

\subsection{Single-cluster case}

\begin{figure} %JO
%PP \begin{figure}[ht]
\epsfxsize=\fsizeb\epsffile{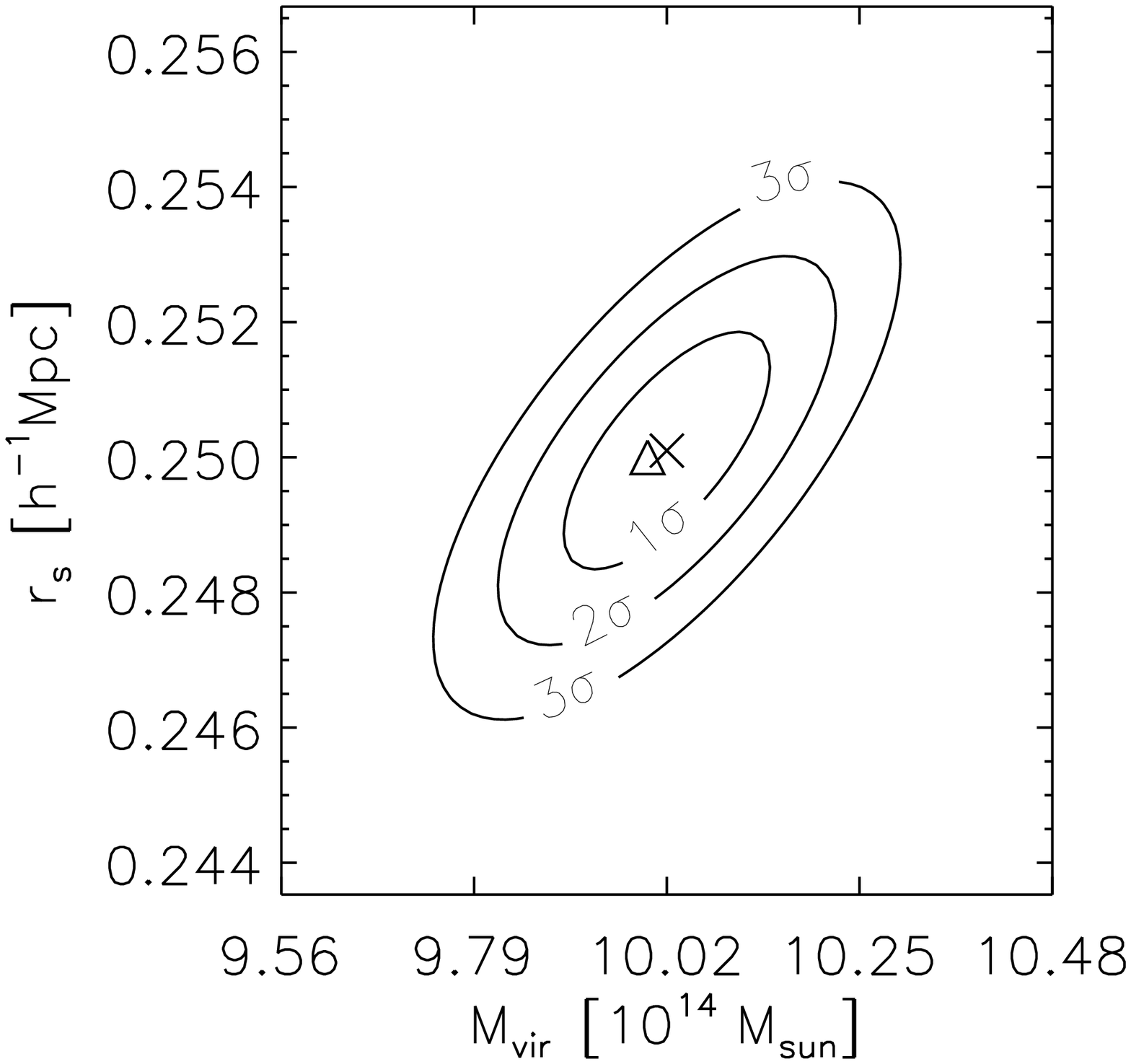} %JO
\epsfxsize=\fsizeb\epsffile{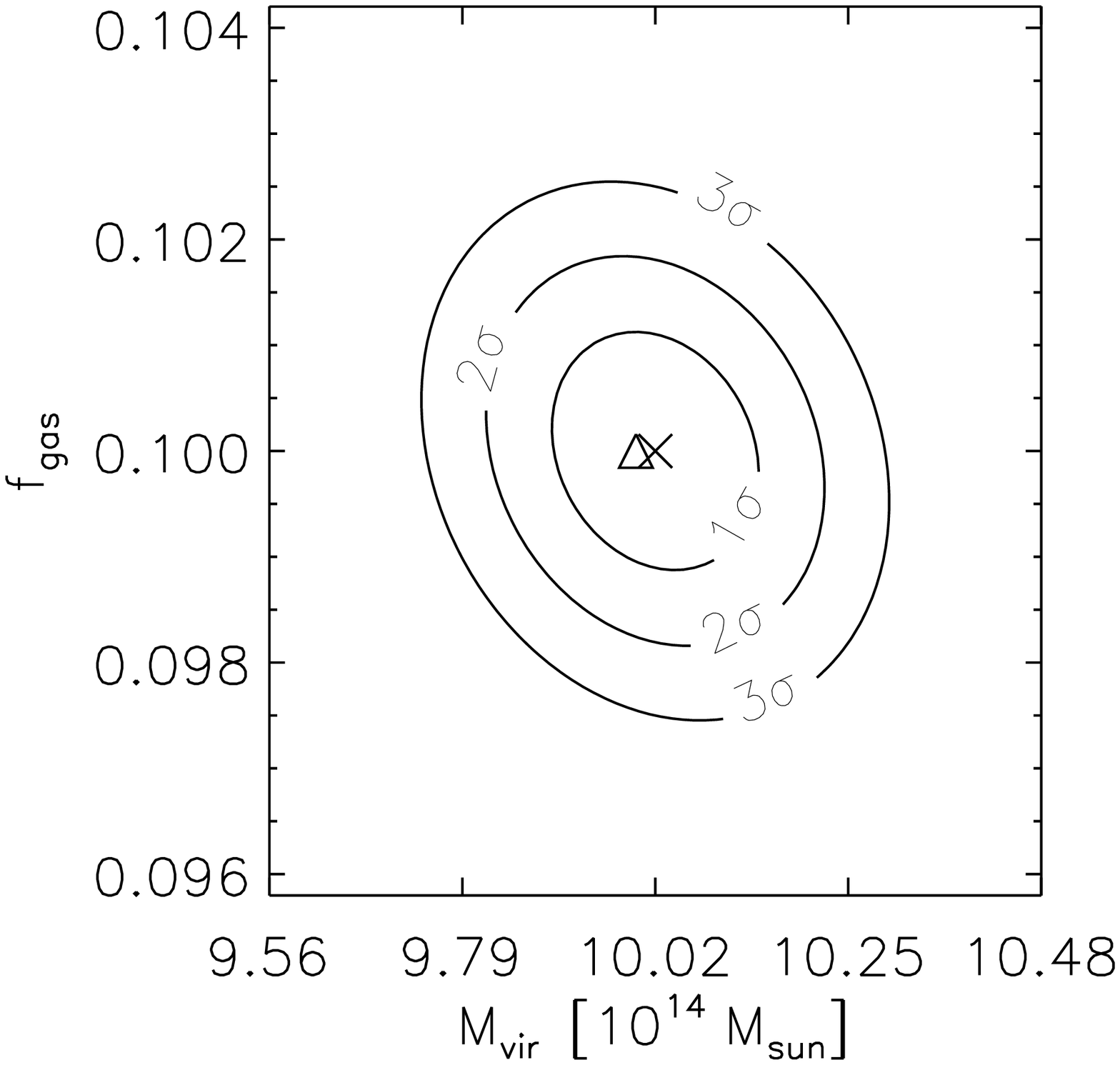} %JO
%PP \mbox{\epsfxsize=\fsizeb\epsffile{fig2a.eps}
%PP \epsfxsize=\fsizeb\epsffile{fig2b.eps}}
\caption{Cuts through the parameter space for fitting synthetic
  observations simulated with one cluster. The parameters are the
  virial mass $M_{\rm vir}$, the scale radius $r_{\rm s}$, and the gas
  fraction within the virial radius $f_{\rm gas}$. Upper panel:
  $\chi^2$ contours in the $M_{\rm vir}$--$r_{\rm s}$ plane; lower
  panel: $\chi^2$ contours in the $M_{\rm vir}$--$f_{\rm gas}$
  plane. The contours are $1\,\sigma$, $2\,\sigma$, and $3\,\sigma$
  confidence levels. The cross marks the best-fit parameters, the
  triangle the parameters of the input model.}
\label{fig:2}
\end{figure}

We begin with ``observations'' created from a single cluster, and try
to fit them with a model of the same functional form as used in the
simulation, consisting of a single cluster. The best-fit parameters
and the goodness-of-fit $G$ at the minimum $\chi^2$ are given in
Tab.~\ref{tab:1}. The $\chi^2$ per degree of freedom in this case is
$1.001$ which, for the number of degrees of freedom we have ($N_{\rm
dof}=2\times1024+121-3=2166$), yields a goodness-of-fit of $G=48.4\%$.
Examples with different realizations of the synthetic data show that
these results are typical.  As the table shows, the input cluster
parameters are well reproduced. $\chi^2$ contours in the $M_{\rm
vir}$--$r_{\rm s}$ and the $M_{\rm vir}$--$f_{\rm gas}$ planes are
shown in Fig.~\ref{fig:2}.

\begin{table} %JO
%PP \begin{table}[ht]
\caption{Parameters obtained from fitting a single cluster with a
  single cluster. $M_{\rm vir}$ is the virial mass, $r_{\rm s}$ is the
  scale radius, $f_{\rm gas}$ is the gas fraction, and $G$ is the
  goodness-of-fit according to eq.~(\ref{eq:gof}). $3\,\sigma$ errors
  are given. On the whole, the input parameters are well recovered,
  and lie all within the $1\,\sigma$ contour level of the
  minimization.}
\label{tab:1}
%PP \medskip
\begin{center}
\begin{tabular}{l*{4}{c}}
\hline
& \multicolumn{3}{c}{Parameter} & \\
\cline{2-4}
& $M_{\rm vir}$ & $r_{\rm s}$ & $f_{\rm gas}$ & $G$ \\
& [$10^{14}\,M_\odot$] & [$h^{-1}\,$Mpc] & [\%] & [\%] \\
\hline
input & $10.0$ & $0.250$ & $10.0$ & -- \\
best fit & $10.0\pm0.3$ & $0.250\pm0.004$ & $10.0\pm0.3$ & $48.4$ \\
\hline
\end{tabular}
\end{center}
\end{table}

\subsection{Double-cluster case}

\begin{table} %JO
%PP \begin{table}[ht]
\caption{Results from attempts at fitting with one cluster synthetic
  data that were simulated with two clusters. The input models consist
  of two clusters whose scale radii $r_{{\rm
  s},1,2}=0.25\,h^{-1}\,$Mpc, gas fraction $f_{\rm gas}=10\%$, and
  total mass $M_{{\rm vir},1}+M_{{\rm vir},2}=10^{15}\,M_\odot$ are
  kept constant, while their mass ratio $m=M_{{\rm vir},1}/M_{{\rm
  vir},2}$ is varied. The table shows the mass $M_{\rm vir}$, scale
  radius $r_{\rm s}$, and gas fraction $f_{\rm gas}$ of the
  best-fitting single-cluster model. The $\chi^2$ is to be compared to
  the number of degrees of freedom, $N_{\rm dof}=2166$. $G$ is the
  goodness-of-fit according to eq.~(\ref{eq:gof}). $3\,\sigma$ errors
  are given.}
\label{tab:2a}
%PP \medskip
\begin{center}
\begin{tabular}{@{\extracolsep\fill}l*{5}{c}@{\extracolsep\fill}}
\hline
    & \multicolumn{3}{c}{Parameter} & & \\
\cline{2-4}
$m$ & $M_{\rm vir}$ & $r_{\rm s}$ & $f_{\rm gas}$ & $\chi^2$ & $G$ \\
    & [$10^{14}\,M_\odot$] & [$h^{-1}\,$Mpc] & [\%] & & [\%] \\
\hline
1:1 & $13.0\pm0.2$ & $0.275\pm0.002$ & $6.8\pm0.1$ & $3129$ &
$<10^{-4}$ \\
1:2 & $12.9\pm0.1$ & $0.271\pm0.001$ & $7.0\pm0.1$ & $2868$ &
$<10^{-4}$ \\
1:3 & $12.6\pm0.1$ & $0.268\pm0.001$ & $7.4\pm0.1$ & $2634$ &
$<10^{-4}$ \\
1:4 & $12.4\pm0.2$ & $0.266\pm0.003$ & $7.7\pm0.2$ & $2504$ &
$<10^{-4}$ \\
1:5 & $11.7\pm0.1$ & $0.265\pm0.001$ & $8.3\pm0.1$ & $2481$ &
$2\times10^{-4}$ \\
1:6 & $11.5\pm0.2$ & $0.263\pm0.002$ & $8.4\pm0.2$ & $2425$ &
$7\times10^{-3}$ \\
\hline
\end{tabular}
\end{center}
\end{table}

We now proceed to ``observations'' simulated with two clusters, with
virial masses $M_{{\rm vir,1}}=(1-m)\,(M_{{\rm vir},1}+M_{{\rm
vir},2})$ and $M_{{\rm vir,2}}=m\,(M_{{\rm vir},1}+M_{{\rm vir},2})$,
projected onto each other along the line-of-sight. The first question
is whether it is possible to significantly detect that the input
cluster is double. This is the case if an attempt to fit the data with
one cluster only results in a best-fit $\chi^2$ which yields an
unacceptable goodness-of-fit. We create synthetic data with two
clusters, keeping the scale radii $r_{\rm s}=0.25\,h^{-1}\,$Mpc, the
gas fraction $f_{\rm gas}=10\%$, and the total cluster mass $M_{{\rm
vir},1}+M_{{\rm vir},2}=10^{15}\,M_\odot$ constant. We then vary the
mass ratio $m=M_{{\rm vir},1}/M_{{\rm vir},2}$. We investigate the
cases $m=\{$1:1, 1:2,$\ldots$, 1:6$\}$.  Typical results are
summarized in Tab.~\ref{tab:2a}. The table shows that for all mass
ratios, the single-cluster models fail to interpret the data
acceptably. In turn, this implies that we can significantly
distinguish between single- and double-cluster cases even if the mass
ratio is fairly large.

\begin{figure} %JO
%PP \begin{figure}[ht]
\epsfxsize=\fsizeb\epsffile{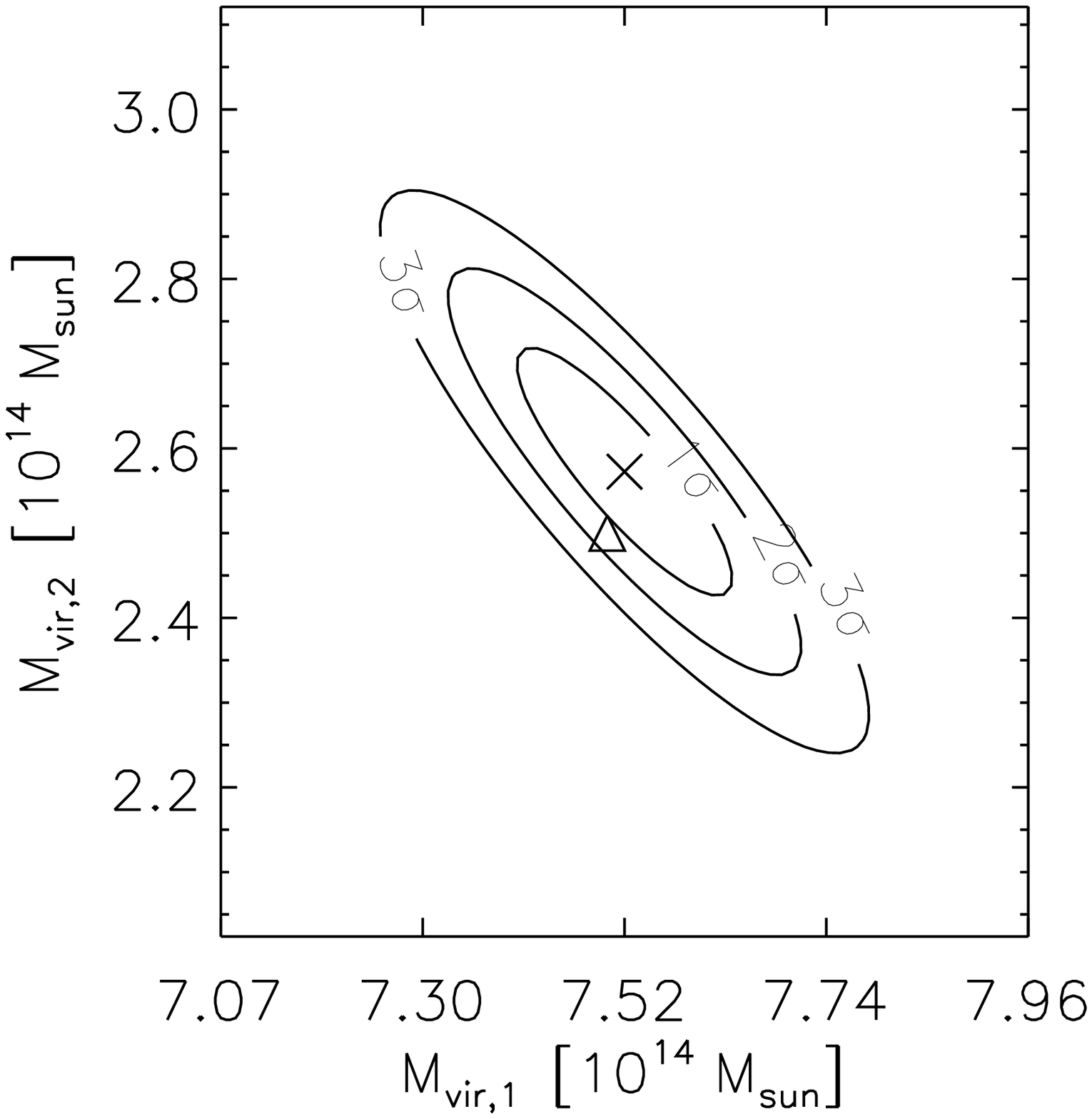} %JO
\epsfxsize=\fsizeb\epsffile{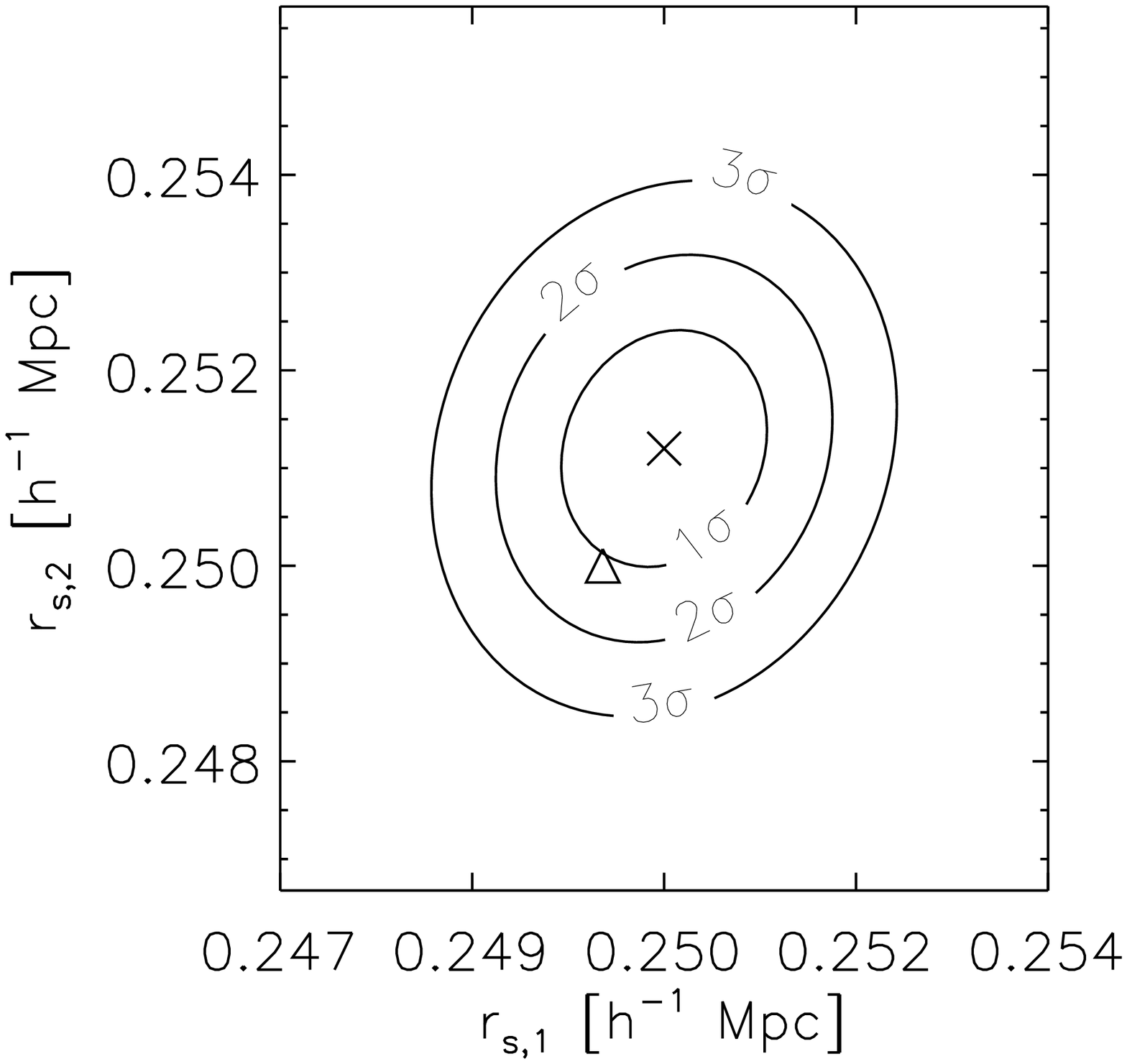} %JO
%PP \mbox{\epsfxsize=\fsizeb\epsffile{fig3a.eps}
%PP \epsfxsize=\fsizeb\epsffile{fig3b.eps}}
\caption{Two cuts through the parameter space for fitting synthetic
  observations simulated with two clusters. There are five parameters
  in total; the two virial masses $M_{{\rm vir},1,2}$, the two scale
  radii $r_{{\rm s},1,2}$, and the gas fraction within the virial
  radius $f_{\rm gas}$ (assumed to be the same in both clusters).  The
  mass ratio is $m=\hbox{1:3}$.  Upper panel: $\chi^2$ contours in the
  $M_{{\rm vir},1}$--$M_{{\rm vir},2}$ plane; lower panel: $\chi^2$
  contours in the $r_{{\rm s},1}$--$r_{{\rm s},2}$ plane. The contours
  are for the same confidence levels as in Fig.~\ref{fig:2}. Crosses
  mark the best-fit parameters, triangles the input parameters.}
\label{fig:3}
\end{figure}

Since for large $N_{\rm dof}$, the $\chi^2$ distribution approaches a
Gaussian with mean $N_{\rm dof}=2166$ and variance
$\sigma_{\chi^2}=(2\,N_{\rm dof})^{1/2}\approx65.8$, the formal
significance for rejecting the single-cluster hypothesis is
$\sim15\,\sigma_{\chi^2}$ for $m=\hbox{1:1}$, and
$\sim4\,\sigma_{\chi^2}$ for $m=\hbox{1:6}$.

\begin{table*} %JO
%PP \begin{table}[ht]
\caption{Results from fitting with a double-cluster model synthetic
  data that were simulated with two clusters. $m$ is the mass ratio,
  $M_{{\rm vir},1,2}$ are the virial masses, $r_{{\rm s},1,2}$ the
  scale radii, $f_{\rm gas}$ is the gas fraction (assumed to be the
  same in both clusters), and $G$ is the goodness-of-fit from
  eq.~(\ref{eq:gof}). $3\,\sigma$ errors are given. The input
  parameters are all well recovered. The errors are somewhat larger
  than in the single-cluster case.}
\label{tab:2}
%PP \medskip
\begin{center}
\begin{tabular}{@{\extracolsep\fill}l*{7}{c}}
\hline
& & \multicolumn{5}{c}{Parameter} & \\
\cline{3-6}
& $m$ & $M_{{\rm vir},1}$ & $M_{{\rm vir},2}$ & $r_{{\rm s},1}$ &
$r_{{\rm s},2}$ & $f_{\rm gas}$ & $G$ \\
& & \multicolumn{2}{c}{[$10^{14}\,M_\odot$]}
& \multicolumn{2}{c}{[$h^{-1}\,$Mpc]} & [\%] & [\%] \\
\hline
input & 1:1 & $5.0$ & $5.0$ & $0.25$ & $0.25$ & $10.0$ & -- \\
best fit & $1.0\pm0.4$ & $5.0\pm1.4$ & $5.0\pm1.4$
         & $0.25\pm0.02$ & $0.25\pm0.02$ & $10.1\pm0.3$
         & $23.1$ \\
\hline
input & 1:2 & $6.7$ & $3.3$ & $0.25$ & $0.25$ & $10.0$ & -- \\
best fit & $0.49\pm0.09$ & $6.7\pm0.6$ & $3.3\pm0.5$
         & $0.25\pm0.01$ & $0.25\pm0.01$ & $10.0\pm0.3$
         & $26.0$ \\
\hline
input & 1:3 & $7.5$ & $2.5$ & $0.25$ & $0.25$ & $10.0$ & -- \\
best fit & $0.33\pm0.04$ & $7.5\pm0.3$ & $2.6\pm0.3$
         & $0.251\pm0.002$ & $0.251\pm0.003$ & $10.0\pm0.2$
         & $24.1$ \\
\hline
input & 1:4 & $8.0$ & $2.0$ & $0.25$ & $0.25$ & $10.0$ & -- \\
best fit & $0.26\pm0.03$ & $8.0\pm0.3$ & $2.1\pm0.2$
         & $0.251\pm0.003$ & $0.250\pm0.004$ & $10.0\pm0.2$
         & $24.5$ \\
\hline
input & 1:5 & $8.3$ & $1.7$ & $0.25$ & $0.25$ & $10.0$ & -- \\
best fit & $0.22\pm0.02$ & $8.4\pm0.2$ & $1.8\pm0.2$
         & $0.251\pm0.002$ & $0.250\pm0.002$ & $10.0\pm0.2$
         & $27.9$ \\
\hline
input & 1:6 & $8.6$ & $1.4$ & $0.25$ & $0.25$ & $10.0$ & -- \\
best fit & $0.17\pm0.01$ & $8.6\pm0.1$ & $1.5\pm0.1$
         & $0.251\pm0.002$ & $0.249\pm0.002$ & $10.0\pm0.2$
         & $27.9$ \\
\hline
\end{tabular}
\end{center}
\end{table*} %JO
%PP \end{table}

The expected trend is noticed, as for $m\to0$ the minimization should
converge to the single cluster result. By further examination of the
figures in the table we notice that the total mass of the system is
always overestimated by about $10-30\%$, the scale radius is always
overestimated by $5-10\%$, and the gas fraction is always
underestimated by $15-30\%$ (for these values of $m$). The quoted
errors do not thus represent the true errors, because the systematic
errors from assuming the wrong model are ignored.  The interpretation
of this deviation is as follows: the minimization routine ``detects''
too low a temperature for the amount of flux it ``sees''. It therefore
tries to increase the amount of flux without changing the temperature,
by widening the potential well (\ie increasing $r_{\rm s}$) without
making the well substantially deeper or equivalently without creating
an unacceptable mismatch with the lensing distortion. This in turn
ends up in attributing higher enclosed mass and a bit too high flux
rate. The cure for the latter is achieved by the reduction of $f_{\rm
gas}$.  This explains the false parameters we get out of the
minimization.

Having seen that we can significantly reject the hypothesis that the
synthetic data were simulated with a single cluster, we should now ask
whether we get an acceptable GoF for the double-cluster model. And
furthermore, how well can we recover the parameters of the individual
clusters? For that purpose, we use the same double-cluster data that
were created earlier once again, and fit them with a double-cluster
model. Table~\ref{tab:2} summarizes the results. The number of degrees
of freedom is now reduced by two, $N_{\rm dof}=2164$. The values of
$\chi^2/N_{\rm dof}$ are now typically $\sim1.02$, resulting in
goodness-of-fit values of $G\sim25\%$. The input parameters are all
well recovered. The $3\,\sigma$ errors are somewhat larger than in the
case of one cluster. They are largest for mass ratio $m=\hbox{1:1}$,
namely $\sim28\%$ for $M_{\rm vir}$, $\sim8\%$ for $r_{\rm s}$, and
$\sim3\%$ for $f_{\rm gas}$, and they decrease to a few percent for
smaller mass ratios. For examples, we show in Fig.~\ref{fig:3} two
cuts through the parameter space of a double-cluster model with mass
ratio $m=\hbox{1:3}$. The upper panel shows $\chi^2$ contours in the
$M_{{\rm vir},1}$--$M_{{\rm vir},2}$ plane, the lower panel shows
$\chi^2$ contours in the $r_{{\rm s},1}$--$r_{{\rm s},2}$ plane. The
fairly large elongation of the $\chi^2$ ellipses in the former case
illustrates the comparatively large uncertainty in the masses: within
a fairly broad range, it is possible to increase or decrease one mass
at the expense of the other.

We also ran simulations where the clusters had different scale radii,
$r_{{\rm s},1}\ne r_{{\rm s},2}$. When the less massive cluster has a
smaller scale radius, it is recovered less precisely, because it is
masked by the larger $r_{\rm s}$ of the dominant, more massive
cluster. Even then, the masses of the individual clusters are well
recovered.

\section{A Comparison to $\beta$ Fits}
\label{sec:comp_beta}

\begin{figure} %JO
%PP \begin{figure}[ht]
\epsfxsize=\fsizeb\leavevmode\epsffile{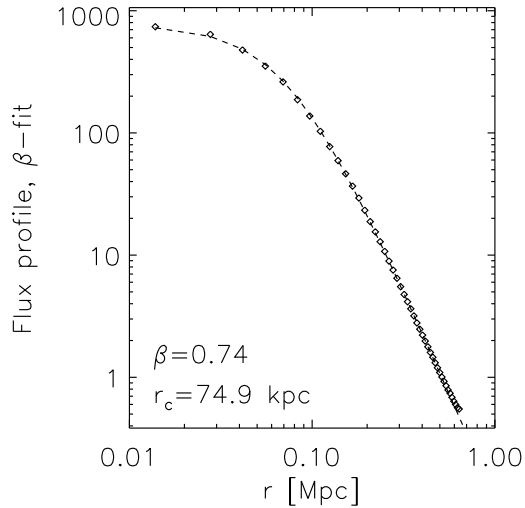} %JO
%PP \begin{center}
%PP \mbox{\epsfxsize=\fsizeb\epsffile{fig4.eps}}
%PP \end{center}
\caption{The azimuthally averaged flux profile (diamonds) of the
  cluster shown in Fig.~\protect\ref{fig:1}, overlaid with a $\beta$
  fit profile (\cf eq.~\protect\ref{eq:6.1}; dashed line). The
  parameters of the fit (the core radius $r_{\rm c}$ and $\beta$) are
  given in the figure. At $r_{\rm lim}=0.49\,h^{-1}\,$Mpc, the flux
  profile drops below the background level. The gas fraction within
  this radius, chosen such that the total cluster X-ray luminosity is
  reproduced, is $f_{\rm gas}'=(18\pm1)\,\%$. The total mass within
  $r_{\rm lim}$, implied by the $\beta$-fit parameters together with
  the temperature given in Fig.~\protect\ref{fig:1} and
  eq.~(\protect\ref{eq:6.2}), is $M_\beta(r_{\rm
  lim})=(2.0\pm0.2)\times10^{14}\,M_\odot$ ($3\,\sigma$ errors).}
\label{fig:4}
\end{figure}

The conventional way to interpret X-ray observations is to azimuthally
average the flux map and fit the functional form
\begin{equation}
  S_{\rm X}(r) \propto \left[
  1+\left(\frac{r}{r_{\rm c}}\right)^2\right]^{-3\beta/2+1/2}
\label{eq:6.1}
\end{equation}
to the resulting flux profile. Assuming that the X-ray emitting gas is
isothermal and in hydrostatic equilibrium with a spherically symmetric
gravitational potential, the total mass implied by eq.~(\ref{eq:6.1})
is
\begin{equation}
  M_\beta(r) = \frac{3\beta\,r\,kT}{G\bar m}\,\frac{x^2}{1+x^2}\;,
\label{eq:6.2}
\end{equation}
where $x=r/r_{\rm c}$. We apply this technique to the flux map shown
in Fig.~\ref{fig:1}. The model (\ref{eq:6.1}) provides an excellent
fit to the flux profile, with $\beta=0.74$ and $r_{\rm
c}=74.9\,h^{-1}\,$kpc (\cf Fig.~\ref{fig:4}). At $r_{\rm
lim}=0.49\,h^{-1}\,$Mpc, the flux profile drops below the background
noise. At that radius, the spectral temperature of
$T=(5.1\pm0.4)\;$keV yields, together with the other parameters,
$M_\beta(r_{\rm lim})=(2.0\pm0.2)\times10^{14}\,M_\odot$ ($3\,\sigma$
errors). Given these results, we can further determine the gas
fraction required to explain the total X-ray flux. At $3\,\sigma$
confidence (only noise included), it turns out to be $f'_{\rm
gas}=(18\pm1)\%$.

The flux map in Fig.~\ref{fig:1} was simulated using {\em two\/}
clusters of $M_{{\rm vir},1}=5\times10^{14}\,M_\odot=M_{{\rm vir},2}$
and $r_{\rm s}=0.25\,h^{-1}\,$Mpc, so that the total mass within the
radius accessible to X-ray observations should be $M_{\rm
total}(r_{\rm lim})=4.26\times10^{14}\,M_\odot$. The input gas
fraction was $f_{\rm gas}=10\%$. Of course, $f'_{\rm gas}$ from the
$\beta$ fit is now the gas fraction within the observable radius
$r_{\rm lim}$ rather than the virial radius $r_{\rm vir}$, with
$r_{\rm lim}\sim r_{\rm vir}/3$, hence the prime on $f_{\rm gas}$. The
gas fraction of the input model slightly depends on $r$. At $r_{\rm
lim}$, it is $f'_{\rm gas}=9.2\%$ rather than $10\%$. $f'_{\rm gas}$
as obtained from the $\beta$-fit technique therefore overestimates the
true gas fraction by a factor of $\sim1.8-2.1$. Obviously, the
estimates from the $\beta$ fit differ substantially from the true
values {\em despite\/} the $\beta$ fit's being excellent and the gas
being in hydrostatic equilibrium within each of the two clusters.

Part of the gross discrepancy between the true gas fraction and that
inferred from the $\beta$ fit comes from the fact that fitting the
X-ray observations alone does not give any clue as to the structure of
the cluster along the line-of-sight. Other contributions emerge from
the attempt to rely on the X-ray data alone without appealing to the
lensing data.

For a single-cluster model with $M_{\rm vir}=10^{15}\,M_\odot$ instead
of the two-cluster model with the same total mass and a mass ratio of
$m=\hbox{1:1}$, the $\beta$-fit mass is $M_\beta(r_{\rm
lim})=(3.6\pm0.4)\times10^{14}\,M_\odot$, while the true mass within
$r_{\rm lim}$ is $3.7\times10^{14}\,M_\odot$. Therefore, in the
single-cluster case, the $\beta$-fit mass is fairly accurate within
the observable radius. The gross overestimate of the gas fraction in
the double-cluster case is thus largely caused by the underestimate of
the total mass.

In addition, the $\beta$ fit profile (\ref{eq:6.1}) implies that the
gas and dark matter density profiles flatten off at radii smaller than
the core radius $r_{\rm c}$. The true (input) dark matter density
profile has a central cusp $\propto r^{-1}$. The isothermal gas
density profile (\ref{eq:5}) approaches the center exponentially,
$\propto\exp(-Ax)$. The actual central gas density is higher than
deduced from (\ref{eq:6.1}), and therefore the actual total gas mass
required to reproduce the observed X-ray flux is smaller than
inferred. Even if we use the $\beta$ model to interpret
``observations'' simulated with only one cluster, the best-fit gas
fractions obtained are still systematically too high by $\sim20-40\%$.

We also performed the counter experiment of simulating data with a
King -- (\ie $\beta=1$) rather than the NFW profile, and fitting them
with the NFW profile.  In this case, the $\beta$-fit technique
recovers the input parameters very well, including the gas
fraction. When the core radius is small, $r_{\rm
c}\lesssim0.2\,h^{-1}\;$Mpc, the fit with the NFW profile fails.  It
yields a marginally acceptable goodness-of-fit of $G=8\%$ when the
core radius is larger, $r_{\rm c}\gtrsim0.25\,h^{-1}\;$Mpc. However,
the best-fit mass is then $M_{\rm
vir}=(11.4\pm0.5)\times10^{14}\,M_\odot$ instead of the input $M_{\rm
vir}=10^{15}\,M_\odot$.

\section{Summary and outlook}
\label{sec:conc}

Resolving the l.o.s.\ density profile of what appears to be a single
cluster is not a hopeless task. In this paper, we have presented an
approach that may ultimately lead to a clear distinction between
different l.o.s.\ profiles. The key idea is to combine all the
available information for the cluster, using simultaneously the X-ray
data (their spatial and energy distribution), and the gravitational
lensing properties of the cluster(s).

Two additional pieces of information were left out of the calculation.
The first is the redshift distribution of cluster galaxies, because of
the possible complication due to velocity bias that may interfere with
the mass estimate, contamination by non-member galaxies along the
line-of-sight, and triple-valued zones. We precluded this information
from the analysis even though it may help demonstrate the existence of
a bimodal distribution in the case of two clusters. The second piece
of information is the distribution of background galaxy sizes. This
was left out because the intrinsic size distribution is broader than
the ellipticity distribution, and consequently the additional
constraints gained from including magnification effects are fairly
weak.

We restricted our investigation to the question of how well a single
cluster can be distinguished from two well separated clusters along
the l.o.s. We have demonstrated, by using realistic simulations of
cluster observations, that a single-cluster model for two clusters
along the l.o.s.\ can be rejected, using the method, on a
$\sim4-15\,\sigma$ level, depending on the mass ratio between the two
clusters.

The true (input) parameters of the system, i.e.\ the total mass
$M_{\rm vir}$ within a certain overdensity level, the scale radius
$r_{\rm s}$, and the gas content $f_{\rm gas}$, can be recovered with
typical ($3\,\sigma$) fractional accuracies of a few percent for all
parameters of single clusters. In the double-cluster case, the errors
are largest when the mass ratio is close to unity, and they decrease
to a few percent for smaller mass ratios. There is no good control
over the separation between the clusters in the case of a two-cluster
system.

We have further shown how wrong results for the cluster parameters can
be obtained by using the $\beta$-fit that allegedly provides an
appropriate fit for the X-ray flux data. Most of this effect can be
ascribed to the fact that the $\beta$-fit technique is unable to
recognize whether an apparent cluster is single or double.  The method
we propose in this paper does not suffer from that drawback, and hence
we propose it supersedes the $\beta$-fit for mass and gas-fraction
estimates.

We note that there is an increasing number of clusters for which there
is evidence that they consist of two projected clusters rather than a
of a single one. A well-known example for this is Abell 1689. Our
choice of the singular NFW density profile is well motivated by
numerical simulations (Navarro \etal 1996; Cole \& Lacey 1996; Huss
\etal 1997), and by observations which demonstrate that the mass
profile derived from galaxy velocity data does not flatten off at
small radii (Carlberg \etal 1997). In addition, strong gravitational
lensing requires cluster cores to be much smaller than inferred from
X-ray observations alone, if cores exist at all.

Reality spans a much broader range than what we examined in this
paper. To begin with, two clusters along the l.o.s.\ can be in the
process of merging. In that case hydrostatic equilibrium ceases to be
a reasonable assumption and so does the isothermal model. Shocks due
to the merging process heat the intracluster medium in an
inhomogeneous fashion that is difficult to model. Hydrodynamic
simulations of clusters may be useful in modeling the shock layer and
its effect on the various X-ray observations.

Then, even for an isolated cluster which does not experience any
merging, there may exist cooling flows that invalidate the assumption
of isothermality (for the validity of hydrostatic equilibrium and
isothermality in the presence of cooling flows, see \eg Waxman \&
Miralda-Escud\'e 1995). There is hope these can be actually observed
and may be azimuthally averaged over in order to regain the ability to
model the cluster.

Another disturbing caveat may lie in the spherical symmetry we assume.
Even though X-ray observations usually are circular on the sky for
clusters (and not elliptical), this may be partially attributed to
selection effects. Optically defined clusters show much more
pronounced two-dimensional elliptical shapes in the galaxy
distribution (\eg Plionis, Barrow, \& Frenk 1991). Although the
ellipticity of the potential is smaller than that of the mass
distribution, some of it must remain. A natural generalization of the
current work would be to investigate families of elliptical potentials
(Bartelmann \& Kolatt 1997).  By introducing the axis ratio as one of
the free parameters, there is a continuous transition between
spherical symmetry and elongated elliptical cluster when both are
assumed to be in a hydrostatic equilibrium. Using the same $\chi^2$
statistic as we used here, with this additional free parameter, should
result in an estimate for the cluster elongation.  In addition, mass
models obtained from large arcs in some clusters can help constrain
the morphology of the projected cluster mass distribution.

The removal of some of the projection effects by the means presented
in this paper will allow a better understanding of the cluster
environment and a safer use of clusters as large scale structure
probes. These are two big leaps forward.

\section*{Acknowledgments}

We thank M.\ Freyberg and S.\ Schindler for providing information on
the ROSAT detectors, C.\ Canizares for advice, and Peter Schneider for
a critical reading of the manuscript.  This work was supported in part
by the Sonderforschungsbereich 375 of the Deutsche
Forschungsgemeinschaft (MB) and the US National Science Foundation
(PHY-9507695; TK).


\begin{thebibliography}{}

\bibitem{ref:1} Abell, G.O., 1958, ApJS, 3, 211

\bibitem{ref:2} Abell, G.O., Corwin Jr., H.G., Olowin, R.P., 1989,
ApJS, 70, 1

\bibitem{ref:3} Bahcall, N.A., 1988, ARA\&A, 26, 631

\bibitem{ref:4} Bahcall, N.A., Cen, R.Y., 1993, ApJ, 407, L49

\bibitem{ref:5} Bartelmann, M., 1996, A\&A, 313, 697

\bibitem{ref:6} Bartelmann, M., Steinmetz, M., 1996, MNRAS, 283, 431

\bibitem{ref:7} Bartelmann, M., Kolatt, T.S., 1997, in preparation

\bibitem{ref:8} Brainerd, T.G., Blandford, R.D., Smail, I., 1996, ApJ,
466, 623

\bibitem{ref:9} Burns, J.O., Ledlow, M.J., Loken, C., Klypin, A.,
Voges, W., Bryan, G.L., Norman, M.L., White, R.A., 1996, ApJ, 467, L49

\bibitem{ref:10} Carlberg, R.G., Yee, H.K.C., Ellingson, E., Morris,
S.L., Abraham, R., Gravel, P., Pritchet, C.J., Smecker-Hane, T.,
Hartwick, F.D.A., Hesser, J.E., Hutchings, J.B., Oke, J.B., 1997,
preprint astro-ph/9703107

\bibitem{ref:11} Cavaliere, A., Fusco-Femiano, R., 1976, A\&A, 49, 137

\bibitem{ref:12} Cole, S., Lacey, C., 1996, MNRAS, 281, 716

\bibitem{ref:13} Dalton, G.B., Efstathiou, G., Maddox, S.J.,
Sutherland, W.J., 1994, MNRAS, 269, 151

\bibitem{ref:14} Ebeling, H., Voges, W., B\"ohringer, H., Edge, A.C.,
Huchra, J.P., Briel, U.G., 1996, MNRAS, 281, 799

\bibitem{ref:15} Eke, V.R., Cole, S., Frenk, C.S., 1996, MNRAS, 282,
263

\bibitem{ref:16} Fadda, D., Girardi, M., Giuricin, G., Mardirossian,
F., Mezzetti, M., 1996, ApJ, 473, 670

\bibitem{ref:17} Gioia, I.M., Maccacaro, T., Schild, R.E., Wolter, A.,
Stocke, J.T., Morris, S.L., Henry, J.P., 1990, ApJS, 72, 567

\bibitem{ref:18} van Haarlem, M.P., Frenk, C.S., White, S.D.M., 1997,
preprint astro-ph/9701103, MNRAS, in press

\bibitem{ref:19} Holzapfel, W.L., Arnaud, M., Ade, P.A.R., Church,
S.E., Fischer, M.L., Mauskopf, P.D., Rephaeli, Y., Wilbanks, T.M.,
Lange, A.E., 1997, preprint astro-ph/9702224

\bibitem{ref:20} Huss, A., Jain, B., Steinmetz, M., 1997, preprint
astro-ph/9703014

\bibitem{ref:21} Lilly, S.J., Tresse, L., Hammer, F., Crampton, D.,
Le F\`evre, O., 1995, ApJ, 455, 108

\bibitem{ref:22} Loeb, A., Mao, S., 1994, ApJ, 435, L109

\bibitem{ref:23} Mazure, A., Katgert, P., den Hartog, R., Biviano, A.,
Dubath, P., Escalera, E., Focardi, P., Gerbal, D., Giuricin, G.,
Jones, B., Le F\`evre, O., Moles, M., Perea, J., Rhee, G., 1996, A\&A,
310, 31

\bibitem{ref:24} Miralda-Escud\'e, J., 1991, ApJ, 370, 1

\bibitem{ref:25} Miralda-Escud\'e, J., Babul, A., 1995, ApJ, 449, 18

\bibitem{ref:26} Navarro, J.F., Frenk, C.S., White, S.D.M., 1996, ApJ,
462, 563

\bibitem{ref:27} Plionis, M., Barrow, J.D., \& Frenk, C.S. 1991,
MNRAS, 249, 662

\bibitem{ref:28} Rephaeli, Y., 1995, ApJ, 445, 33

\bibitem{ref:29} Roettiger, K., Stone, J.M., Mushotzky, R., 1997,
preprint astro-ph/9702072

\bibitem{ref:30} Schneider, P., Seitz, C., 1995, A\&A, 294, 411

\bibitem{ref:31} Snowden, S.L., Freyberg, M.J., Plucinsky, P.P.,
Schmitt, J.H.M.M., Tr\"umper, J., Voges, W., Edgar, R.J., McCammon,
D., Sanders, W.T., 1995, ApJ, 454, 643

\bibitem{ref:32} Sunyaev, R.A. \& Zel'dovich, Ya.B. 1980, ARA\&A, 18,
237

\bibitem{ref:33} Tanaka, Y., Inoue, H., Holt, S.S., 1994, PASJ, 46,
L37

\bibitem{ref:34} Tyson, J.A., Seitzer, P., 1988, ApJ, 335, 552

\bibitem{ref:35} Viana, P.T.P., Liddle, A.R., 1996, MNRAS, 281, 323

\bibitem{ref:36} Waxman, E., Miralda-Escud\'e, J., 1995, ApJ, 451, 451

\bibitem{ref:37} White, S.D.M., Efstathiou, G., Frenk, C.S., 1993,
MNRAS, 262, 1023

\bibitem{ref:38} Zwicky, F., Herzog, E., Wild, P., Karpowicz, M.,
Kowal, C.T., 1968, Calif. Inst. for Technology, Pasadena

\end{thebibliography}
\end{document}